\documentclass[%
preprint,
superscriptaddress,
 amsmath,amssymb,
 aps,
floatfix
]{revtex4-2}

\usepackage{graphicx}
\usepackage{dcolumn}
\usepackage{bm}
\usepackage[ngerman,english,]{babel}
\usepackage[usenames,dvipsnames]{xcolor}
\usepackage{siunitx}
\usepackage{appendix}

\begin{document}


\title{Automated matching of two-time X-ray photon correlation maps from protein dynamics with Cahn-Hilliard type simulations using autoencoder networks}

\author{Sonja Timmermann}
\email{sonja.timmermann@uni-siegen.de}
\affiliation{Department Physik, Universit\"at Siegen, Walter-Flex-Str. 3, 57072 Siegen, Germany}

\author{Vladimir Starostin}
\email{vladimir.starostin@uni-tuebingen.de}
\affiliation{Institut für Angewandte Physik, Universit\"at T\"ubingen, Auf der Morgenstelle 10, 72076 T\"ubingen, Germany}%

\author{Anita Girelli}
\affiliation{Institut für Angewandte Physik, Universit\"at T\"ubingen, Auf der Morgenstelle 10, 72076 T\"ubingen, Germany}%

\author{Anastasia Ragulskaya}
\affiliation{Institut für Angewandte Physik, Universit\"at T\"ubingen, Auf der Morgenstelle 10, 72076 T\"ubingen, Germany}%

\author{Hendrik Rahmann}
\affiliation{Department Physik, Universit\"at Siegen, Walter-Flex-Str. 3, 57072 Siegen, Germany}

\author{Mario Reiser}
\affiliation{Stockholm University, SE-106 91 Stockholm, Sweden}

\author{Nafisa Begam}
\affiliation{Institut für Angewandte Physik, Universit\"at T\"ubingen, Auf der Morgenstelle 10, 72076 T\"ubingen, Germany}%

\author{Lisa Randolph}
\affiliation{Department Physik, Universit\"at Siegen, Walter-Flex-Str. 3, 57072 Siegen, Germany}

\author{Michael Sprung}
\author{Fabian Westermeier}
\affiliation{Deutsches Elektronen-Synchrotron DESY, Notkestraße 85, 22607 Hamburg, Germany}
 
\author{Fajun Zhang}
\author{Frank Schreiber}
\email{frank.schreiber@uni-tuebingen.de}
\affiliation{Institut für Angewandte Physik, Universit\"at T\"ubingen, Auf der Morgenstelle 10, 72076 T\"ubingen, Germany}%

\author{Christian Gutt}
\email{christian.gutt@uni-siegen.de}
\affiliation{Department Physik, Universit\"at Siegen, Walter-Flex-Str. 3, 57072 Siegen, Germany}

\date{\today}

\begin{abstract}
We use machine learning methods for an automated classification of experimental XPCS two-time correlation functions from an arrested liquid-liquid phase separation of a protein solution. We couple simulations based on the Cahn-Hilliard equation with a glass transition scenario and classify the measured correlation maps automatically according to quench depth and critical concentration at a glass/gel transition. 
We introduce routines and methodologies using an autoencoder network and a differential evolution based algorithm for classification of the measured correlation functions. The here presented method is a first step towards handling large amounts of dynamic data measured at high brilliance synchrotron and X-ray free-electron laser sources facilitating fast comparison to phase field models of phase separation.
\end{abstract}

\maketitle


\section{\label{sec:Intro}Introduction}
Phase separation is an ubiquitous process in nature with applications and consequences for a wide range of scientific disciplines such as solid state physics, material sciences and biology. Phase separation in biological systems attracted considerable attention with the discovery that liquid-liquid phase separation (LLPS) in protein solutions constitutes a possible pathway for organizing membraneless structures in living cells \cite{brangwynneGermlineGranulesAre2009,shinLiquidPhaseCondensation2017,berryPhysicalPrinciplesIntracellular2018}. Detailed investigations were carried out with respect to the biological functions of these protein condensates, encompassing biochemical reaction rates, buffering protein concentrations, as well as sensing or signaling \cite{shinLiquidPhaseCondensation2017}.
A variety of diseases which result from a loss/and or change of function of these condensates are accompanied by phase transitions \cite{malinovskaProteinDisorderPrion2013,weberGettingRNAProtein2012}. 

Phase separation is initiated by quenching a system into the metastable region of the phase diagram launching a process of out-of-equilibrium self-organization \cite{dongReincarnationsPhaseSeparation2021}. The state of the condensates depends on the dynamic and kinetic processes during their formation with dynamical asymmetries between the two phases on a hierarchy of length and time scales and invoking viscoelastic properties of the resulting network structures \cite{berryPhysicalPrinciplesIntracellular2018,zaccarelliColloidalGelsEquilibrium2007,tanakaViscoelasticPhaseSeparation2000}. In the biological context an understanding of the comprised time scales is of importance as they are expected to meet the intrinsic time scales of biochemical processes in the condensates. 

Upon phase separation, the dynamics slow down on molecular length scales due to local concentration changes. This microscopic deaccceleration can eventually lead to the arrest of phase separation on larger length scales complemented by the formation of bicontinuous gel network structures – as observed in colloidal and protein systems \cite{manleyGlasslikeArrestSpinodal2005,conradArrestedFluidfluidPhase2010,luGelationParticlesShortrange2008}. Employing time resolved scattering experiments the kinetics during  arrested phase separations have been observed frequently as a slowdown of the growth of static structure factor S(Q) both in Q-position and peak height for low temperature quenches 
\cite{gibaudCloserLookArrested2009,gibaudPhaseSeparationDynamical2011,cardinauxInterplaySpinodalDecomposition2007,davelaKineticsLiquidLiquid2016,davelaArrestedTemporarilyArrested2017,bucciarelliUnusualDynamicsConcentration2015,davelaInterplayGlassFormation2020}. 
In contrast, the dynamics of protein solutions evolving into an arrested phase transition is still elusive as it requires the concurrent monitoring of an extraordinarily broad range of time and length scales. 
Likewise, experimental validation of models of the dynamics of critical phenomena during LLPS like the Cahn-Hilliard equation (CHE) \cite{cahnPhaseSeparationSpinodal1965} and other correlated models is still due, particularly with respect to glass-gel transitions, displaying considerable dynamical asymmetries between the dilute and concentrated phase \cite{berryPhysicalPrinciplesIntracellular2018,girelliMicroscopicDynamicsUnderlying2020}. We note that the CHE has already been used in different machine learning tasks previously  \cite{wightSolvingAllenCahnCahnHilliard2020,pokuriInterpretableDeepLearning2019,zhangMachineLearningMaterials2020,farimaniDeepLearningPhase2018}.

An experimental method capable of accessing the dynamics of protein systems over length scales that range from nano- to micrometer and time scales from microseconds to hours is X-ray photon correlation spectroscopy (XPCS) \cite{luHowLiquidBecomes2008,zhangDynamicScalingColloidal2017,madsenStructuralDynamicsMaterials2016,perakis2020towards}. In XPCS, time series of coherent X-ray speckle patterns are measured giving access to dynamical properties via time resolved correlation maps - the two time correlation function (TTC). Due to the fast megapixel 2D X-ray detectors XPCS experiments can produce large amounts of data in a short time interval with up to thousands of TTCs per hour of beamtime. Given the typical duration of a synchrotron experiment of a few days, the resulting amounts of TTCs are difficult to handle. Therefore, there is a generic need in methods facilitating fast and reliable analysis including classification of the data. This is especially important with regards of steering, selecting and controlling the experimental parameters during the beamtime. However, also in the aftermath of the experiment quick classification methods are important for benchmarking models which for example model a gel transition or the solidification process of a protein solution upon LLPS. 

Machine learning (ML) has already proven to be useful in the analysis of various structural X-ray data from small angle X-ray scattering (SAXS) \cite{archibaldClassifyingAnalyzingSmallangle2020,chenMachineLearningDeciphers2020,frankeMachineLearningMethods2018,wangXrayScatteringImage2016} and wide angle X-ray scattering (WAXS) \cite{chenMachineLearningDeciphers2020, wangXrayScatteringImage2016}, diffraction \cite{berntsonApplicationNeuralNetwork2003,oviedoFastInterpretableClassification2019,vecseiNeuralNetworkbasedClassification2019} and reflectometry \cite{grecoFastFittingReflectivity2019} experiments. The algorithms provide real time feedback \cite{wangXrayScatteringImage2016,keConvolutionalNeuralNetworkbased2018,grecoFastFittingReflectivity2019} during the experiment or sort out bad images to reduce the stored data volume \cite{wangXrayScatteringImage2016,keConvolutionalNeuralNetworkbased2018}. In many cases the machine learning algorithms were trained with synthetical structural data generated from existing data bases on proteins \cite{frankeMachineLearningMethods2018,archibaldClassifyingAnalyzingSmallangle2020}, RNA \cite{chenMachineLearningDeciphers2020} or crystal structures \cite{oviedoFastInterpretableClassification2019,vecseiNeuralNetworkbasedClassification2019}, which enables them to classify the results obtained in experiments without an expert spending time on labeling large amounts of data for training \cite{berntsonApplicationNeuralNetwork2003}. 

The focus of X-ray scattering applications to ML has been mainly on static properties up to now. We note, however, that many processes in nature, such as the LLPS of protein solutions, display dynamic phenomena which require to develop schemes and methods capable of classifying the corresponding dynamic X-ray signatures of the processes. Here, we report on a first step towards this goal and present a neural network method that allows to analyze and classify dynamical X-ray information during a LLPS. We employ the CHE in a simplified 2D version and couple it with a concentration induced glass/gel transition as proposed by \cite{sciortinoInterferencePhaseSeparation1993,jackleGeneralisedHydrodynamicsLangevin1990} and aim for a ML based classification of the experimental data from \cite{girelliMicroscopicDynamicsUnderlying2020}. We aim for an automated assignment of the measured TTCs with the simulated TTCs using the two simulations parameters quench depth $\epsilon$ (i.e. temperature below the critical point) and $\Psi_{gel}$, i.e. the concentration at which the mobility drops by a factor of 1/2. We train an autoencoder neural network with TTC simulation data and employ a differential evolution based algorithm for matching encoded experimental TTCs with simulations.  
Our work constitutes a first step for making use of the static, kinetic and dynamical information contained in XPCS data for ML based analysis of processes in phase separation. 

\section{Analysis and Results}
\subsection{Simulation of the Cahn-Hilliard equation}

\begin{figure*}
\includegraphics[width=\textwidth]{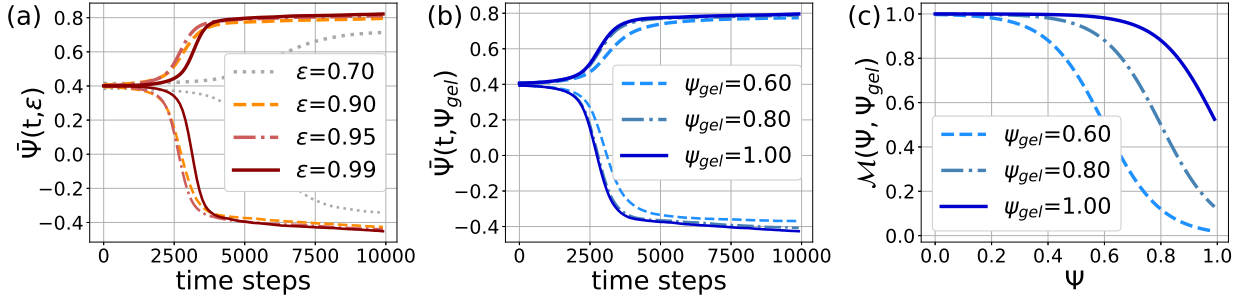}
\caption{\label{fig:MobAndConcentration}(a),(b) Order parameter averaged over all lattice sites having an order parameter higher/lower than the initial order parameter ($\Psi=0.4$) as a function of time. In a), $\Psi_{gel}$ was fixed to $\Psi_{gel}=1$ and $\epsilon$ was varied. For $\epsilon=0.7$, the phase separation is much slower and the concentration of the dense phase is lower at the end of the chosen time window. This makes it hard to see between the effects of high $\Psi_{gel}$, which is why $\epsilon$ is restricted to $[0.88,0.99]$ for the generation of the simulated training data. In (b), $\epsilon$ was fixed to $\epsilon=0.9$ and $\Psi_{gel}$ was varied. The phase separation is starting later for larger values of $\epsilon$ due to the smaller initial fluctuations. For constant $\epsilon$, the phase separations show similar behavior at early times but the droplet formation is faster for larger values of $\Psi_{gel}$. (c) Mobility as a function of the order parameter and the gelation point $\Psi_{gel}$ marking the order parameter at which the mobility drops to $0.5$.}
\end{figure*}

We model the dynamic processes during the spinodal decomposition with the help of the Cahn-Hilliard equation (CHE) \cite{sciortinoInterferencePhaseSeparation1993,guntonDynamicsFirstOrder1983},
\begin{equation}
\frac{\partial \Psi(\boldsymbol{r},t)}{\partial t} = \nabla \mathcal{M} \nabla \left[\Psi^3(\boldsymbol{r},t)-\epsilon\Psi(\boldsymbol{r},t)-\nabla^2 \Psi(\boldsymbol{r},t) \right],
\label{eq::CahnHilliard}
\end{equation}
where $\Psi(\boldsymbol{r},t)$ is the order parameter at given time $t$ and position $\boldsymbol{r}$ which is related to the local protein concentration $\Phi$ by $\Phi=(\Psi+1)/2$ \cite{sciortinoInterferencePhaseSeparation1993}. The symmetry breaking is introduced by the parameter $\epsilon$ that can be identified as being proportional to the reduced quench depth $\epsilon \propto (T_c - T)/T_c$, where $T_c$ is the upper critical temperature of the system. If $\epsilon$ fulfils the condition $\epsilon > 3\Psi^2$ (e.g. \cite{guntonDynamicsFirstOrder1983}), the system is in an unstable state below the spinodal curve and separates into regions with higher and  lower local protein concentrations (Fig. \ref{fig:MobAndConcentration}(a),(b)).

The phase separation by spinodal decomposition is coupled to a glass/gel transition by introducing a concentration dependent mobility parameter $\mathcal{M}$ in eq. \eqref{eq::CahnHilliard} \cite{sappeltComputerSimulationStudy1997} which reduces the mobility of the highly concentrated phase via:
\begin{equation}
\mathcal{M}(\Psi)=\frac{1}{1+\exp(\alpha\cdot(\Psi-\Psi_{gel}))}.
\label{eq::mobility}
\end{equation}
In this model the mobility decreases from values $\mathcal{M}=1$ to $\mathcal{M}=\frac{1}{2}$ when the order parameter $\Psi$ exceeds the value of $\Psi_{gel}$ (Fig. \ref{fig:MobAndConcentration}(c)). The parameter $\alpha$ is fixed to $\alpha=10$ providing a steep decrease of the mobility as required for an arrested phase separation \cite{sappeltComputerSimulationStudy1997}.\\

We perform the simulation on a two dimensional grid of size $256\times 256$. The grid uses periodic boundary conditions and is initialized with a mean order parameter of $\Psi_0=0.4$ such that we observe droplets with a low concentration instead of an interconnected network which would be the case for $\Psi_0=0$. The fluctuations necessary for LLPS are added as a random noise with a maximum amplitude of $0.5\times(1-\epsilon)$. 10.000 time steps are calculated for each simulation with a fixed time step of $\Delta t=0.03$. The temporal evolution of the order parameter for different parameter configurations is depicted in Fig. \ref{fig:MobAndConcentration}(a),(b). Apart from the initial noise configuration, the simulations differ only in the values of the parameters $\epsilon$ and $\Psi_{gel}$. Both were systematically varied within the intervals 
\begin{align}
    \epsilon \in [0.88,0.99] \mbox{ and }
    \Psi_{gel} \in [0.55,1].
    \label{al:parameterRanges}
\end{align}
These intervals were chosen with $\Psi_{gel}>\Psi_0$, thus avoiding an arrest of the phase separation before the formation of droplets. We restrict our analysis to large values of $\epsilon$ as for small values of $\epsilon$ the concentration of the dense phase is too low (gray line in Fig \ref{fig:MobAndConcentration}(a)) to introduce visible changes in the simulations for different values of $\Psi_{gel}$.

\begin{figure}
\includegraphics[width=0.49\textwidth]{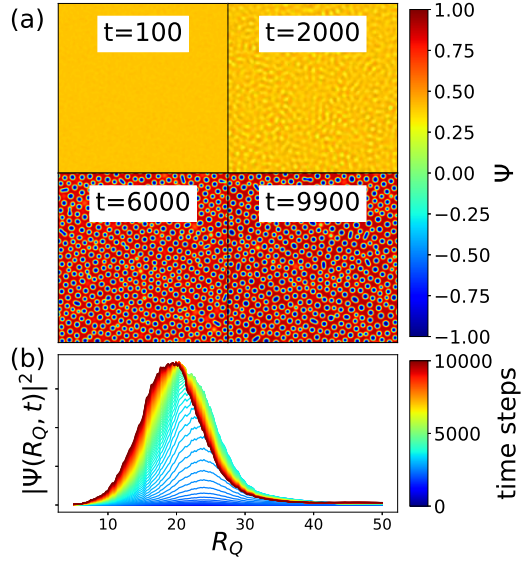}
\caption{\label{fig:RealSpaceAndQProfile}(a) Evolution of the real space configuration of the sample on the $256\times 256$ grid with simulation parameters $\Psi_{gel}=0.6$ and $\epsilon=0.955$, showing the creation and coarsening of droplets with a low concentration. (b) Corresponding temporal evolution of the azimuthally integrated scattering intensity as a function of the absolute value of the distance $R_Q$ to the origin of the Fourier map. The peak in the intensity profile is shifting towards smaller $R_Q$ with time indicating the coarsening of the droplets.}
\end{figure}

A single simulation run yields a time series of the protein density's real space configuration $\Psi(\boldsymbol{r},t)$ (Fig. \ref{fig:RealSpaceAndQProfile}(a)) which is converted to an X-ray speckle pattern by means of Fourier transform yielding a $256\times 256$ image. The temporal evolution of the square of the azimuthally integration of these Fourier images $|\Psi(R_Q,t)|^2$ is shown in Fig. \ref{fig:RealSpaceAndQProfile}(b) and is equivalent to the X-ray intensity. We introduce $R_Q$ as the distance from the center of the Fourier image in unit of pixels. In Fig. \ref{fig:RealSpaceAndQProfile}(b) the occurrence of a peak indicates the phase separation.

The dynamics on a certain length scale in Fourier space can be traced by correlating pixelwise the corresponding intensities with a two time correlation function (TTC)
\begin{equation}
c^{(2)}(t_1,t_2)=\frac{\left\langle{[I(t_1)-\left\langle{I(t_1)}\right\rangle][I(t_2)-\left\langle{I(t_2)}\right\rangle]}\right\rangle}{[\left\langle{I^2(t_1)}\right\rangle-\left\langle{I(t_1)}\right\rangle^2][\left\langle{I^2(t_2)}\right\rangle-\left\langle{I(t_2)}\right\rangle^2]},
\label{eq::TTCcalculation}
\end{equation}
with $\langle \cdot \rangle$ being the average over pixels within a distance $[R_Q-\mathrm{d}R_Q,R_Q+\mathrm{d}R_Q]$ to the center.

The azimuthally integrated intensity profile (Fig. \ref{fig:RealSpaceAndQProfile}(b)) shows the growth dynamics of the droplets by a shift of the peak position towards smaller values of $R_Q$. These dynamics can be observed in the TTC by choosing a value of $R_Q$ such that it is located at the falling slopes of the peaks in Fig. \ref{fig:RealSpaceAndQProfile}(b) which is $R_Q\in[26,45]$. For our training data, we chose $R_Q$ to be an integer in this range, while $\mathrm{d}R_Q$ increases linearly from $0.7$ at $R_Q = 26$ to $3$ at $R_Q = 45$ as it is done in experimental analyses as well to compensate for lower photon counts at higher $Q$ values.


\subsection{Experimental data}
\begin{figure}
\includegraphics[width=0.49\textwidth]{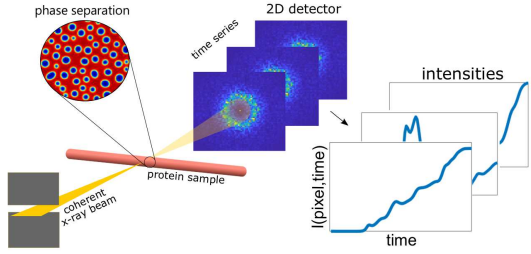}
\caption{\label{fig:expAufbau}Schematic of the experimental setup. The coherent X-ray beam is scattered from the protein solution during the liquid-liquid phase separation yielding a time series of X-ray speckle patterns. A constant Q-ring is selected in the speckle patterns and the intensities are correlated pixel-by-pixel for different times.  }
\end{figure}

The XPCS experiments (Fig. \ref{fig:expAufbau}) were conducted at the Coherence Applications beamline P10 at PETRA III, DESY, employing an X-ray beam of photon energy 8.54 \si{\kilo\electronvolt}, size of 100$\times$100 \si{\micro\meter\squared} and maximum photon density of $10^7$ \si{photons\per\second\per\micro\meter\squared} (for further experimental details see \cite{girelliMicroscopicDynamicsUnderlying2020}). We measured time series of coherent diffraction patterns that were collected with an EIGER 4-megapixel detector covering a $\mathrm{Q}$-range from 3 \si{\per\micro\meter} to 50 \si{\per\micro\meter} . The samples consisted of Immunglobulin G (IgG) with polyethylene glycol (PEG) and NaCl and have been quenched from 37 \si{\degreeCelsius} to six different quench-temperatures below the binodal line. Further details about the sample preparation can be found in the work of Da Vela et al. \cite{davelaArrestedTemporarilyArrested2017}.

The experimental TTCs were calculated similarly to the TTCs from the simulation using Eq. \ref{eq::TTCcalculation}. Regarding the temporal shift of the integrated intensity profiles as a function of $Q$ the falling edge can be restricted to $Q\in [7.5\si{\per\micro\meter},10\si{\per\micro\meter}]$. The experimental TTCs are preprocessed by normalizing on a first off diagonal to address the issue of the low contrast of the measurements.

\subsection{Data Preparation and Autoencoding}

For the training data, 12\,000 pairs of the parameters $\epsilon$ and $\Psi_{gel}$ were sampled from a uniform distribution within the intervals. To increase the simulation robustness against the random initial conditions, TTCs from five simulations with different initial noise were averaged. Thus, $5 \times 12\,000 = 60\,000$ simulations were performed in total. TTCs were calculated for 20 $R_Q$ ranges, so that the training data consisted of $20 \times 12\,000 = 240\,000$ TTCs.

\begin{figure}
\includegraphics[width=0.49\textwidth]{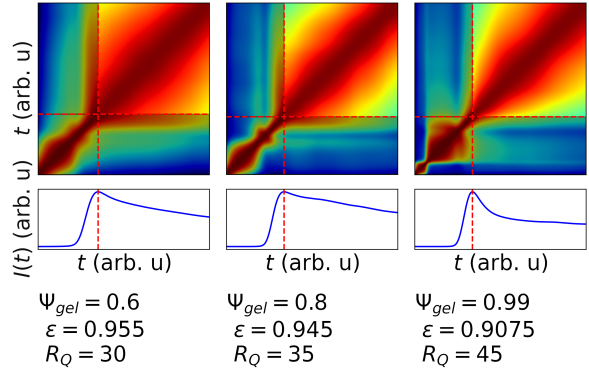}
\caption{A part of the simulated TTCs shows a broad feature at early times (lower left corner of the TTCs). In the second row, the mean intensity of the respective Q range is shown as a function of time. The maximum of this intensity curve defines the point where the TTCs are cropped.}
\label{fig::TTCcutting}
\end{figure}

All simulated TTCs display a broad, slow part in the beginning (Fig. \ref{fig::TTCcutting}) which is absent in the experimental data. This feature is associated with the formation of large clusters in the early phase of the simulation with concentrations close to the initial concentration (Fig. \ref{fig:RealSpaceAndQProfile}(a), t=2000). These clusters lead to a scattering signal at small values of $R_Q$ causing the broad part in the lower left corner of the simulated TTCs. Approaching smaller values of $R_Q$ this feature becomes even more pronounced. An inclusion of thermal fluctuations in the Cahn-Hilliard equation \cite{binderSpinodalDecomposition1980,guntonDynamicsFirstOrder1983} could eventually reduce such features which are not observed in the experiment.

These features are eliminated by removing the part at early times for both experimental and simulated TTC. The time at which the TTCs are cropped is given by the time at which the intensity in the associated Q-ring reaches a maximum. The cutting procedure is illustrated in Fig. \ref{fig::TTCcutting} for three simulated TTCs. Finally, we downscale the resulting TTCs to $64 \times 64$ resolution.

Instead of matching raw TTCs between the experiment and the simulations, we compare their encoded representations via an autoencoder \cite{rumelhart_parallel_1987}. The autoencoder network was trained with the simulated TTCs encoding them in 32-dimensional vectors $\boldsymbol{Z}$. For the encoder, the first 3 layers of the pretrained ResNet-18 model \cite{resnet_2016} were used as a feature extractor, followed by adaptive average pooling with an output size of $3 \times 3$ and a fully-connected network. The decoder architecture follows \cite{deep_consistent_vae_2017}. 

\begin{figure}
\includegraphics[width=0.49\textwidth]{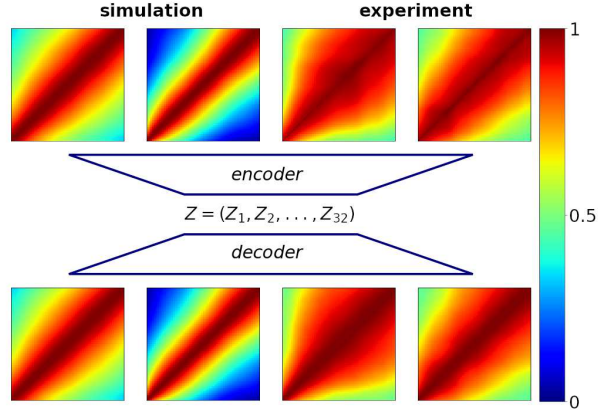}
\caption{Performance of the autoencoder on simulated and experimental TTCs. The encoder network encodes the TTCs into a 32 dimensional vector from which the decoder network restores the TTC. On the experimental data, this procedure removes artifacts of the experimental TTC while keeping the important features.}
\label{fig::AutoEnc2}
\end{figure}

\begin{figure}
\includegraphics[width=0.49\textwidth]{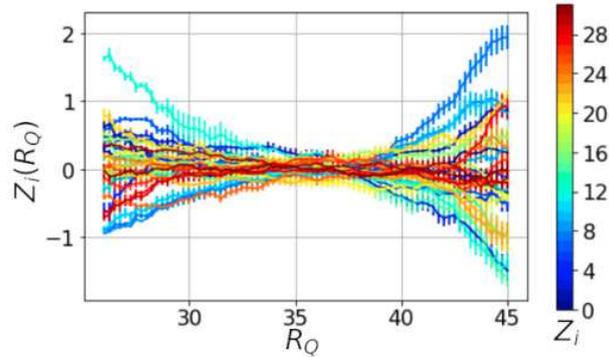}
\caption{Example of encoded vector components $Z_i$ dependence on $R_Q$ for a simulation with $\epsilon=0.88$ and $\Psi_{gel}=0.7$. The error bars result from averaging over 10 simulations with different initial noise.}
\label{fig::AutoEnc1}
\end{figure}

The network is capable of removing experimental artifacts from the experimental TTCs while keeping the important features. Thus, the encoded representations allow comparing TTCs in a more reliable and fast way. The performance of the autoencoder on simulated and experimental data is shown in Fig. \ref{fig::AutoEnc2}. If a series of TTCs from the same simulation but at different values of $R_Q$ was created, the entries of the $\boldsymbol{Z}$ vector change continuously (Fig. \ref{fig::AutoEnc1}) which enabled us to interpolate TTCs for arbitrary values of $R_Q$ in between. 

\subsection{Results}
Our goal is to determine the two model parameters $\epsilon$ and $\Psi_{gel}$ for the experimental measurements performed at different quench temperatures by matching the encoded TTCs from the experiment with encoded TTCs from the simulation. During this process we also aimed to obtain the three additional calibration parameters
\begin{align}
    t_{low},t_{up},k_r,
\end{align}
where $t_{low}$ and $t_{up}$ are the times at which the experimental TTCs are cropped such that they match the timescale of the simulation. $k_r$ denotes the factor for converting the Q-range in units of pixels, that were used in the simulation, to units of inverse micrometer as they were used in the experiment via 
\begin{equation}
    R_Q[\mathrm{pixels}] \cdot k_r = Q[\si{\per\micro\meter}].
\end{equation}
A differential evolution \cite{price_differential_1996} based algorithm was employed for this matching process. For each of the six experimental measurements TTCs were calculated at five different Q values:
\begin{align*}
    Q = 7.02, 7.35, 7.66, 7.96, 8.2 \si{\per\micro\meter}
\end{align*}
such that we obtain a set of 30 experimental TTCs. The differential evolution algorithm optimizes the mean dot product $\xi$ between the normalized encoded representations of 30 experimental TTCs and 30 simulated TTCs from the training dataset with respect to $\epsilon, \Psi_{gel}, k_r, t_{up}, t_{low}$:

\begin{equation}
    \xi = \frac{1}{30} \sum_{q=1}^{5}\sum_{T=1}^{6}Z_{sim}^{q}(\epsilon(T),\Psi_{gel}(T),k_r)\cdot Z_{exp}^{T,q}(t_{up},t_{low}).
\end{equation}

The boundaries for the calibration parameters were chosen based on the assumption that the duration of the experiment is longer than the one of the simulation, and that the chosen $Q$ range $[7.02; 8.2]$ $\si{\per\micro\meter}$ lies within the used $R_Q$ range for the simulations. Knowing that $\epsilon$ is proportional to $(T_c-T)/T_c$, we apply additional boundaries to take into account only those solutions where the average value of $\epsilon$ is decreasing as a function of experimental quench temperature. 

During the fitting procedure the experimental TTCs were cropped and encoded for each generated combination of the parameters. A caching technique was used to accelerate these calculations. The encoded vectors of the simulated TTCs were taken from the training data and interpolated based on the generated calibration coefficient  $k_r$.

\begin{figure*}
\includegraphics[width=\textwidth]{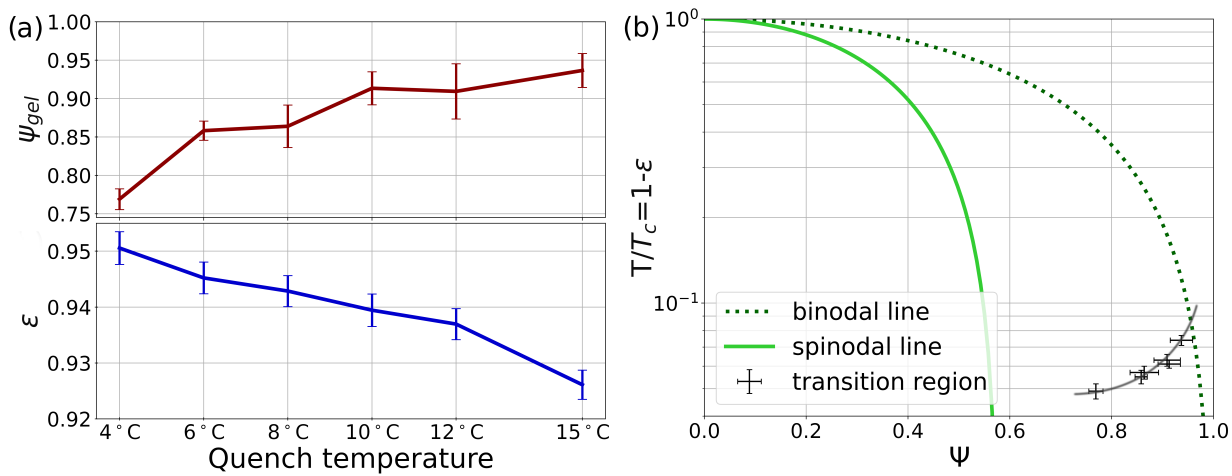}
\caption{\label{fig::Phasediagram}(a) Fitted parameters $\epsilon$ and $\Psi_{gel}$ as a function of quench temperature (b) Phase diagram derived from the free energy density used in the simulation displaying the spinodal and binodal line. The black points represent the predictions of the neural networks for the experimental TTCs. These points mark the region of the glass/gel transition (guide to the eye) in the framework of the simulation.}
\end{figure*}

After 1.500 iterations the result of the fit converged to $\xi\approx 0.72$.
Figure \ref{fig::Phasediagram}(a) shows the averaged outcomes for the simulation parameters $\Psi_{gel}$ and $\epsilon$. The rise in $\Psi_{gel}$ for shallower quenches indicates that the glass transition is occurring at higher concentrations at these temperatures. The ($\Psi_{gel},\epsilon$) pairs were integrated into a phase diagram (Fig. \ref{fig::Phasediagram}(b)) which displays the spinodal and binodal lines as determined from the Landau free energy (Eq. \ref{eq::CahnHilliard}). The predictions of the neural network enable us to estimate the corresponding glass/gel line (grey line as a guide to the eye). This line bends towards the spinodal line resulting in lower concentrations of the dense phase for deeper quenches. Such a behavior of the glass line into the coexistence region has also been observed for the LLPS of $\mathrm{BSA-YCl_3}$ \cite{davelaInterplayGlassFormation2020}, lysozyme \cite{cardinauxInterplaySpinodalDecomposition2007} and for $\gamma$-globulin \cite{davelaArrestedTemporarilyArrested2017}.

Averaging over the 30 outcomes for the calibration parameters yields:
\begin{align}
    k_r &=(0.185\pm0.002)\si{\per\micro\meter}/\mathrm{pixel}, \\
    t_{low}&=(5.02\pm1.72)\si{\second},\\
    t_{up}&=(353.3\pm5.7)\si{\second}.
\end{align}

A further validation of these results with respect to $\epsilon$ and $\Psi_{gel}$ for the experimental TTCs is difficult as both parameters are not directly accessible in the experiment.

\section{Discussion and conclusion}
In conclusion, we successfully trained an autoencoder neural network with TTC data originating from Cahn-Hillard simulations. This autoencoder was used to remove artefacts in TTCs originating from XPCS experiments of LLPS in a model protein solution of IgG. We used a differential evolution based algorithm for matching encoded TTCs from the experiment and the simulation and determine thereby the parameters necessary for connecting time- and lengthscales of simulation to experiment. The simulation parameters of the matching TTCs were used to construct a phase diagram with the experimental data indicating the position of a glass/gel line.

The proper training of neural networks based on dynamical simulation data is challenging. The Cahn-Hilliard model employed here, including the glass transition, is rather simplistic and uses two labeling parameters 
only. However, we emphasize the possibility of extending this methodology to more complicated models capable of  making strong predictions and eventual also bench-marking of theoretical models based on very large XPCS data sets possibly also by including the kinetic information into the model as well. This will be of general interest not only for protein dynamics but also for material science aspects undergoing spinodal decomposition.

\section{Acknowledgements}
S.T. and V.S. contributed equally to this work. S.T. and C.G. acknowledge BMBF for financial support under 05K19PS1 and 05K20PSA. F.Z. acknowledges BMBF for financial support under 05K20VTA. A.R. thanks the Studienstiftung des deutschen Volkes for a personal fellowship. N.B. thanks the Alexander von Humbold Foundation for postdoctoral research fellowship.

\bibliographystyle{apsrev4-2}
\bibliography{refs}

\end{document}